\documentclass{emulateapj}
\usepackage{epsfig} 
\usepackage{amsmath}
\usepackage{amssymb}
\usepackage{natbib}
\usepackage{graphicx}

\shorttitle{Evolution of BH scaling relations in galaxy mergers}
\shortauthors{Johansson et al.}

\begin{document}

\title{The evolution of Black Hole scaling relations in galaxy mergers}

\author{Peter H. Johansson$^1$, Andreas Burkert$^1$, Thorsten Naab$^1$}
\affil{$^1$ Universit\"ats-Sternwarte M\"unchen, Scheinerstr.\ 1, D-81679 M\"unchen, 
Germany; \texttt{pjohan@usm.lmu.de} \\}

\begin{abstract}

We study the evolution of black holes (BHs) on the $M_{\rm BH}-\sigma$
and $M_{{\rm BH}}-M_{\rm bulge}$ planes
as a function of time in disk galaxies undergoing mergers. 
We begin the simulations with the progenitor black hole masses being initially
below $(\Delta \log M_{\rm BH,i}\sim -2)$, on $(\Delta \log M_{\rm BH,i}\sim 0)$ and above
$(\Delta \log M_{\rm BH,i}\sim 0.5)$ the observed local relations. The final
relations are rapidly established after the final coalescence of the galaxies
and their BHs. Progenitors with low initial gas fractions 
($f_{\rm gas}=0.2$) starting below the relations evolve onto the relations $(\Delta \log M_{\rm BH,f}\sim -0.18)$, 
progenitors on the relations stay there $(\Delta \log M_{\rm BH,f}\sim 0)$ and finally progenitors above the
relations evolve towards the relations, but still remaining above
them $(\Delta \log M_{\rm BH,f}\sim
0.35)$. Mergers in which the progenitors have high initial gas fractions
($f_{\rm gas}=0.8$) evolve above the relations in all cases  $(\Delta \log
M_{\rm BH,f}\sim 0.5)$. We find that the initial gas fraction is the prime source of 
scatter in the observed relations, dominating over the scatter arising from
the evolutionary stage of the merger remnants. The fact that BHs starting
above the relations do not evolve onto the relations, indicates that our simulations
rule out the scenario in which overmassive BHs evolve onto the relations through gas-rich mergers.
By implication our simulations thus disfavor the
picture in which supermassive BHs develop significantly before their parent bulges.

\end{abstract}

\keywords{galaxies: active --- galaxies: evolution --- galaxies: interactions --- methods: numerical }

\section{Introduction}

Observations in recent years have revealed a strong correlation in the
local Universe between the central supermassive black holes (BHs) and their host galaxies as manifested
in the relation between the BH mass and the bulge velocity dispersion, 
$M_{\rm BH}-\sigma$
(e.g. \citealp{2000ApJ...539L...9F,2002ApJ...574..740T,2009ApJ...698..198G};), 
the bulge stellar mass $M_{{\rm BH}}-M_{\rm bulge}$ (e.g. \citealp{1998AJ....115.2285M};
\citealp{2004ApJ...604L..89H}), the concentration of light in the galaxy
(e.g. \citealp{2001ApJ...563L..11G}) and the bulge binding energy, $M_{{\rm BH}}-M_{\rm
  bulge}\sigma^{2}$ (e.g. \citealp{2007ApJ...665..120A}). 
The evolution of these relations with redshift is still unclear with some
studies finding evolution in the $M_{\rm BH}-\sigma$ \citep{2006ApJ...645..900W,2008ApJ...681..925W}
and $M_{{\rm BH}}-M_{\rm bulge}$
(\citealp{2006ApJ...649..616P,2007ApJ...667..117T,2009arXiv0911.2988D})
relations, with the high redshift BHs being overmassive
for a fixed $\sigma$ and $M_{\rm bulge}$ compared to
the local relations, whereas other studies are consistent with no redshift evolution in the observed
correlations \citep{2009arXiv0908.0328G,2009ApJ...706L.215J}.
A possible explanation for this discrepancy lies in the uncertainties in observational
selection biases and in the evolution in the intrinsic scatter that is
typically stronger for larger BH masses \citep{2007ApJ...670..249L}. 

The observed correlations are typically explained using theoretical models 
relying on some form  of self-regulated BH
mass growth, in which gas is fed to the central black hole until the black
hole releases sufficient energy to unbind the gas and blow it away in
momentum- or pressure-driven winds (e.g. 
\citealp{1998A&A...331L...1S,1999MNRAS.308L..39F,2001ApJ...554L.151B,2007ApJ...665.1038C,2009ApJ...699...89C}). 
The observed correlations and their evolution with redshift have been
reproduced in semi-analytic models (e.g. \citealp{2006MNRAS.365...11C,2008MNRAS.391..481S}), in self-consistent numerical
simulations of both isolated galaxies and galaxy mergers
(\citealp{2005Natur.433..604D,2005MNRAS.361..776S,2006ApJ...641...90R}) as
well as in galaxies simulated in a full cosmological setting
(\citealp{2007MNRAS.380..877S,2008ApJ...676...33D,2009MNRAS.398...53B}). 
The key assumption in these models is that the galaxies
undergo a brief radiatively-efficient quasar phase triggered by gas-rich
galaxy mergers (e.g. \citealp{2002ApJ...580...73T,2006ApJS..163....1H,2008ApJS..175..356H}) during
which the bulk of the BH growth is taking place and the observed correlations are established.

In a previous paper (\citealp{2009ApJ...690..802J}, hereafter J09), we showed
that the merger remnants of both equal- and unequal-mass mergers of disk
and elliptical galaxies satisfy the observed $M_{\rm BH}-\sigma$ and 
$M_{{\rm BH}}-M_{\rm bulge}$  correlations.
In this Letter, we study for the first time in detail the evolution of the BH scaling relations
during disk galaxy mergers using a new sample of high resolution
simulations including a self-consistent formulation for BH feedback.
We seed the BHs initially with masses corresponding to locations below, on 
and above the observed relations. Thus, in contrast to previous
studies we study here for the first time also overmassive BHs lying initially above 
the observed relations. The BH scaling relations are defined for merger remnants that
have reached their final dynamical state and it is not obvious if the
scaling relations are valid during the merging process and at what stage the
galaxies evolve onto the relations.

\begin{figure}
\centering
\includegraphics[width=7.4cm]{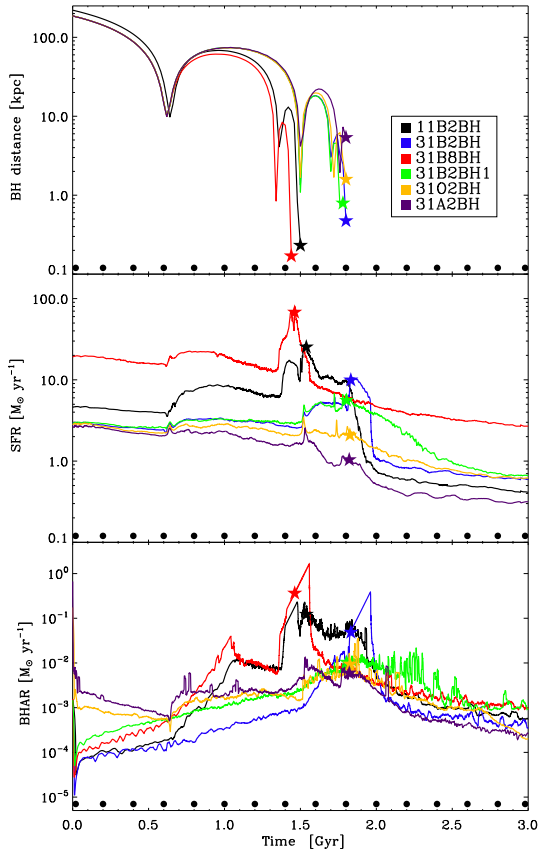}
\caption{The relative distance between the BHs (top), the total SFR (middle), and the
  evolution of the total BH accretion rate (bottom) as a function of time. The star
  symbol indicate the time of BH coalescence and the filled circles at the bottom of
  the plots indicate the time
   at which ($M_{\rm BH}$,$\sigma$,$M_{\rm bulge}$)
  are evaluated in Figs. \ref{MBH_evo_new_BBH}-\ref{MBH_evo_new_ABH}.}
\label{Dist_SFR_BH-evo}
\end{figure}

\begin{table*}
\caption{The simulated merger sample}             
\scriptsize{
\label{sims}      
\centering          
\begin{tabular}{c c c c c c c c c c c c c c}
\hline\hline       
  
Model &  $M_{\rm BH,i1}$\footnote[$a$]{Initial BH mass in $ 10^{5} M_{\odot}$}
&   $M_{\rm BH,i2}$\footnotemark[$a$] & Mass ratio & $\alpha$ & $f_{\rm gas}$ 
& $N_{\rm{gas,tot}}$ & $N_{\rm{disk,tot}}$  & $N_{\rm{bul,tot}}$  &
  $N_{\rm{DM,tot}}$ &  $M_{\rm BH,f}$\footnote[$b$]{Final BH mass in
    $ 10^{5} M_{\odot}$} & $\sigma_{\rm bul,f}$\footnote[$c$]{Final stellar
    velocity dispersion in $\rm km/s$} & $M_{\rm bul,f}$\footnote[$d$]{Final
    bulge mass in $10^{10} M_{\odot}$} \\
\hline                    
11B2BH & 1.0 &  1.0 & 1:1 & 25 & 0.2 & 120 000 & 480 000 & 200 000 & 800 000 & 459  &  183.8  &  5.07  \\  
31B2BH & 1.0 & 1.0 & 3:1 & 25 & 0.2 &  80 000 & 320 000 & 133 333 & 533 333 & 271 & 151.8 & 1.90 \\
31B8BH & 1.0 & 1.0 & 3:1 & 25 & 0.8 &  320 000 & 80 000 & 133 333 & 533 333 &   1136 &  181.6  & 2.97  \\
31B2BH1 & 1.0 & 1.0 & 3:1 & 100 & 0.2 &  80 000 & 320 000 & 133 333 & 533 333  &   82.5 & 144.7 &  1.26   \\
\hline
11O2BH & 159 &  159 & 1:1 & 25 & 0.2 & 120 000 & 480 000 & 200 000 & 800 000  &  600  & 174.6 &   3.38 \\  
31O2BH & 159 & 36.4 & 3:1 & 25 & 0.2 &  80 000 & 320 000 & 133 333 & 533 333  & 253 & 132.0 & 1.55 \\
31O8BH & 159 & 36.4 & 3:1 & 25 & 0.8 &  320 000 & 80 000 & 133 333 & 533 333 &   1698 &  155.2  &  3.02  \\
\hline
11A2BH & 477 & 477 & 1:1 & 25 & 0.2 & 120 000 & 480 000 & 200 000 & 800 000 &   1200 &  164.2 &  3.35 \\ 
31A2BH & 477 & 109 & 3:1 & 25 & 0.2 &  80 000 & 320 000 & 133 333 & 533 333   & 652 & 128.7 & 1.59 \\
31A8BH & 477 & 109 & 3:1 & 25 & 0.8 &  320 000 & 80 000 & 133 333 & 533 333 &    842 & 122.3 & 1.21 \\
\hline                  
\end{tabular}
}
\end{table*}

\begin{figure*}
\centering 
\includegraphics[width=14.9cm]{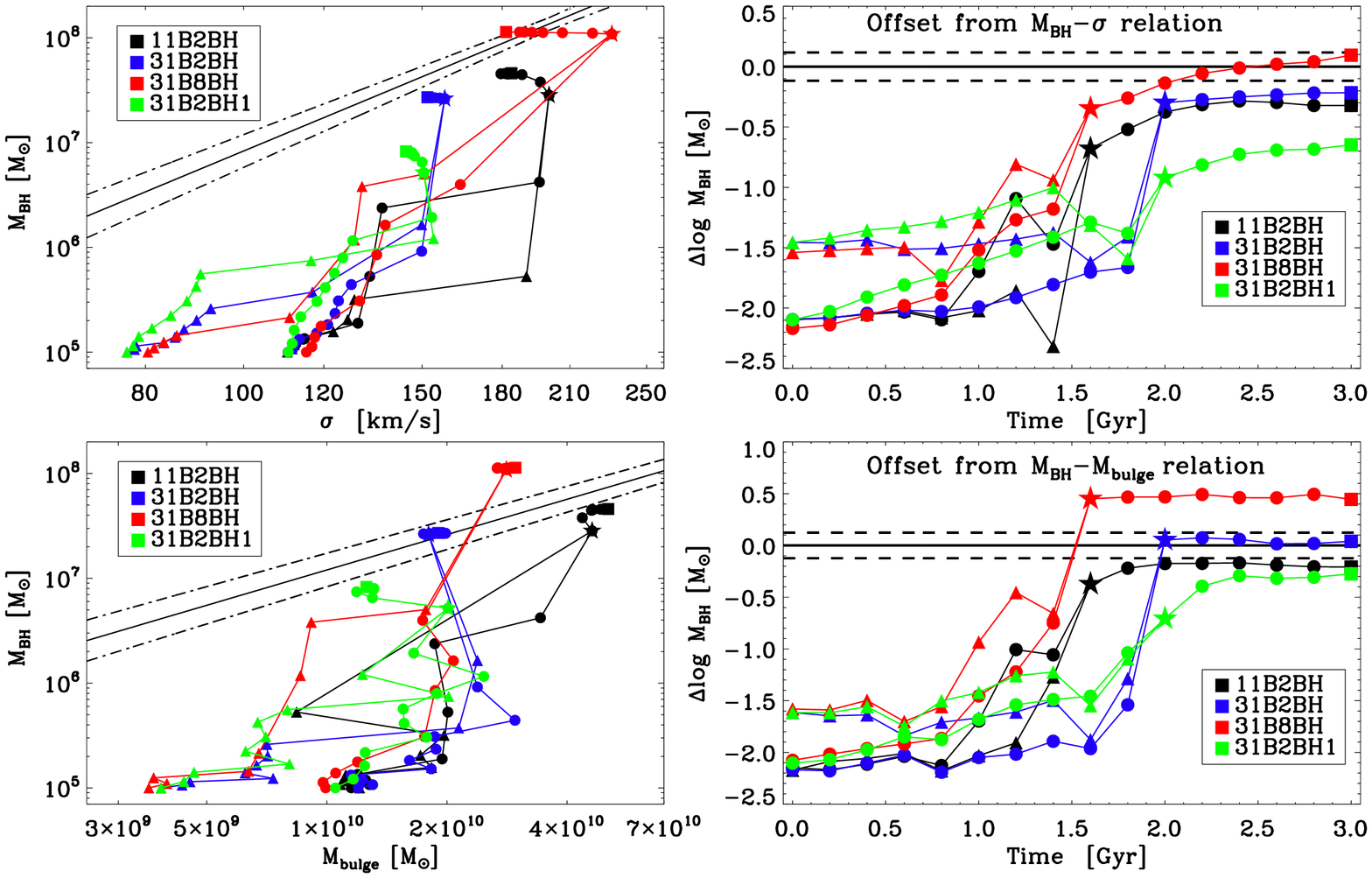}
\caption{The evolution of the BHs in the B-merger sample  $(\Delta \log M_{\rm
    BH,i}\sim -2)$ as a function of time on the $M_{\rm
    BH}-\sigma$ plane (top left panels) overplotted by lines giving the observed
  correlation from \citet{2002ApJ...574..740T}. The bottom left panel gives
  the evolution on the  $M_{\rm BH}-M_{\rm bulge}$ plane overplotted by the observed
  correlation from \citet{2004ApJ...604L..89H}. Each point is separated by $\Delta t=0.2 \
    \rm Gyr$, with filled circles indicating the primary galaxy, triangles indicating
    the secondary galaxy, stars showing the time of BH coalescence and filled squares the final state. 
    The panels on the right give the logarithmic offset
    (Eq. \ref{eq:offset}) from the corresponding observed relation as a function of time.}
\label{MBH_evo_new_BBH}
\end{figure*}

\section{Simulations}

The simulations were performed using the TreeSPH-code GADGET-2
\citep{2005MNRAS.364.1105S} on the local Altix 3700 Bx2 machine.
The code includes radiative cooling for a primordial composition of
hydrogen and helium. Star formation and the
associated supernova feedback is included using the sub-resolution model 
of \citet{2003MNRAS.339..289S}. In this model the ISM is treated as a 
two-phase medium \citep{1977ApJ...218..148M,2006MNRAS.371.1519J} in which cold 
clouds are embedded in a tenuous hot gas at pressure equilibrium.  
BH feedback is modeled using the \citet{2005MNRAS.361..776S} effective model
in which the BH sink particles accrete gas from the surrounding medium according to a 
Bondi-Hoyle-Lyttleton parametrization 
\begin{equation}
\dot{M}_{\rm{B}}=\frac{4 \pi \alpha G^{2} M_{\rm BH}^{2} \rho}{(c_{\rm s}^2+v^{2})^{3/2}},
\label{Bondi}
\end{equation}
 with the maximum accretion rate set by the Eddington limit.
Here $\rho$ and $c_{\rm s}$ are the density and sound speed of the
surrounding gas, respectively, $v$ is the velocity of the BH relative  
to the surrounding gas and $\alpha$ is a dimensionless efficiency parameter.
Following J09 a total of 0.5\% of accreted the rest mass energy is then  
injected as thermal energy into the gas surrounding the BH particle.
Throughout this paper we use the BH repositioning method that ensures 
rapid merging of the BHs once the two BHs reside in the same parent
galaxy. We assume that the BHs merge instantly if they enter the smoothing length of each
other and if their relative velocity is below the local sound speed.

The progenitor galaxies are setup following the method of
\citet{2005MNRAS.361..776S} and as detailed in
J09 with all primary galaxies having $v_{\rm vir}=160 \ \rm km s^{-1}$,$r_{\rm vir}=160 \ h^{-1}\rm{kpc}$
corresponding to a virial mass of 
$M_{\rm vir}=9.53\times 10^{11} h^{-1} M_{\odot}$\footnote{$h=0.71$ is defined
  such that $H_{0}=100h \ \rm{km s^{-1} Mpc^{-1}}$} analogous to the Milky Way.
The dark matter profiles in all model galaxies are described by a \citet{1990ApJ...356..359H}
profile constructed using the concentration parameter $c=9$ of the corresponding NFW halo \citep{1997ApJ...490..493N}.
The dark matter halos are populated with exponential disks with a baryonic mass
fraction of $m_{d}=0.041$ and a stellar Hernquist bulge with a fraction
$m_{b}=0.01367$ $(m_{b}=1/3m_{d})$ of the total virial mass, $M_{\rm vir}$. The disk is setup with a fractional gas content 
of $f_{\rm gas}$, with the rest being disk stars, with the scale lengths of the
disk and bulge computed as detailed in J09. The secondary models in the
unequal-mass mergers are identical to the primary models, except for the fact
that all components are scaled down in mass by a factor of three.

In setting up the primary galaxy disks we use a relatively high
numerical resolution of 300,000 disk particles of which a fraction $f_{\rm gas}$
are gaseous, 100,000 bulge particles and 400,000 dark matter particles. Each
gas particle can only spawn a single stellar particle resulting in a constant
particle mass resolution of $m_{\rm{bar}}=1.30\times 10^{5} h^{-1} M_{\odot}$ for all
baryonic particles and $m_{\rm{DM}}=2.25\times 10^{6} h^{-1} M_{\odot}$
for the dark matter particles. The gravitational softening length was 
set to  $\epsilon=0.02 h^{-1} \rm{kpc}$ for all baryonic particles
and the BHs and to $\epsilon=0.083 h^{-1} \rm{kpc}$ for the dark matter particles.
Finally we insert a seed black hole 
at rest in the center of each galaxy model, where we vary the BH mass
depending on if the models should lie initially below (B-series), on
(O-series) or above (A-series) the observed local BH scaling relations. 

We adopt orbital geometry G13 \citep{2003ApJ...597..893N} for all
simulations in this paper. This geometry
corresponds to the inclinations $i_{p}=-109, i_{s}=180$ and the arguments of the
pericenter $\omega_{p}=60, \omega_{s}=0$ for the primary and secondary
galaxies, respectively. The galaxies approach each other on a parabolic orbit where the initial separation
of the progenitors is $R_{\rm init}=0.5(r_{\rm vir,p}+r_{\rm vir,s})$ 
and the pericentric distance is $r_{\rm peri}=r_{\rm d,p}+r_{\rm d,s}$,
where $r_{\rm vir,p}$, $r_{\rm d,p}$ and $r_{\rm vir,s}$, $r_{\rm d,s}$ are
the virial and disk scale radii for the primary and secondary galaxies, respectively. 
All simulations were evolved for a total of $t=3 \ \rm Gyr$ with the merger
typically taking place at $t\sim 1.5 \ \rm Gyr$. The
main numerical parameters and final properties of all 10 simulations are summarized in Table \ref{sims}.

The parameters governing the multi-phase feedback are as detailed in J09 resulting
in a star formation rate (SFR) of $\sim 1 M_{\odot}\rm yr^{-1}$ for our
primary galaxy. The accretion parameter $\alpha$ (Eq. \ref{Bondi}) can
essentially be seen as an empirical correction factor that translates from 
the resolvable low mean density to the time-averaged small-scale density at 
the location of the BH. We found that for $\alpha=25$ the BHs reach Eddington 
limited mass growth during the merger. This is not the case for the value
$\alpha=100$ used in J09 (see Fig. \ref{Dist_SFR_BH-evo}). The corresponding
evolution on the $M_{\rm BH}-\sigma$ plane is such that the 3B2BH1
$(\alpha=100)$ simulation does not evolve onto the relation, whereas the 
3B2BH $(\alpha=25)$ evolves onto the relations reproducing the lower 
resolution results from J09 (see Fig. \ref{MBH_evo_new_BBH}). We conclude that  as the spatial 
resolution in the simulations is increased the corresponding $\alpha$ value 
needs to be decreased in order to reproduce the lower resolution runs. Thus, we
adopt $\alpha=25$ for all subsequent simulations. Finally, we note that our
results do not depend critically on $\alpha$ as long as the
condition of Eddington limited mass growth during the merger is fulfilled.

\begin{figure*}
\centering 
\includegraphics[width=14.9cm]{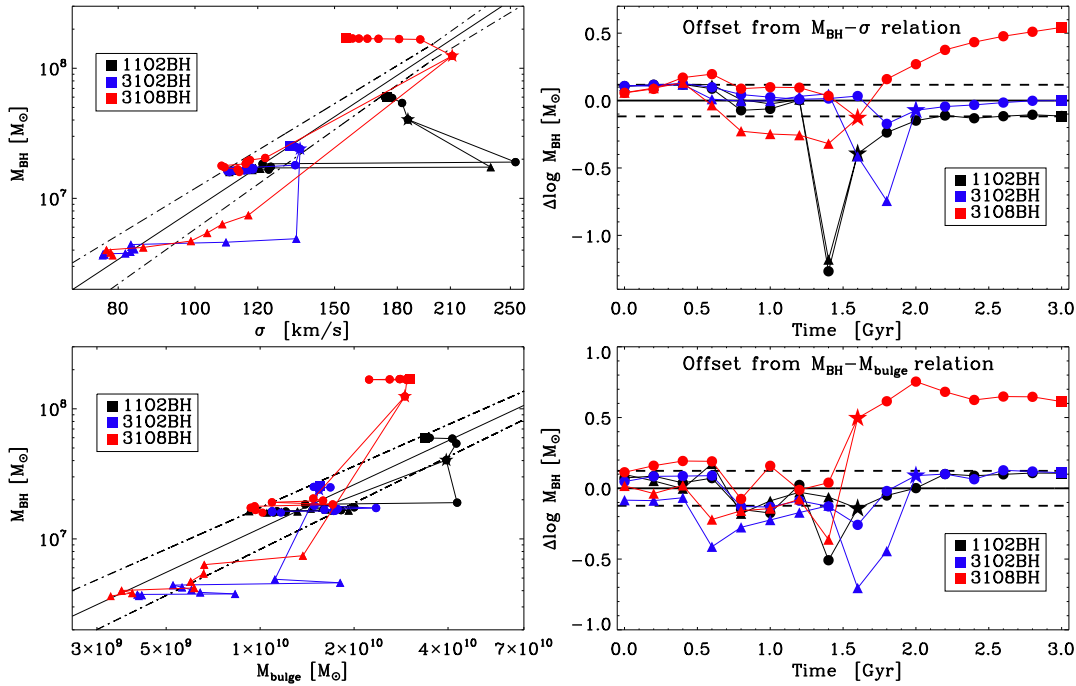}
\caption{The evolution of the O-merger sample with the BHs starting
  initially on the local scaling relations $(\Delta \log M_{\rm
    BH,i}\sim 0)$.  Symbols and the observed relations as
  defined in Fig. \ref{MBH_evo_new_BBH}.}
\label{MBH_evo_new_OBH}
\end{figure*}

\section{Methodology}
\label{methods}

We plot in Fig. \ref{Dist_SFR_BH-evo} the relative distance between the BHs,
the total SFR and the total BH accretion rate for a subsample of six
simulations. In addition, we overplot filled circles at time intervals of
$\Delta t=0.2 \ \rm Gyr$ throughout the 3 Gyr simulation indicating the time
at which the galaxies are extracted for analysis. This gives a total of
16 snapshots during the merger. We perform the analysis using all stellar
particles found in a sphere with a radius of $r=30 \ \rm kpc$ centered on the
BH of the corresponding galaxy. The repositioning technique employed in this 
study ensures that the BHs always trace the highest density central structure of their parent halo.

The inclusion of a massive BH initially has a marked effect on the
star formation and BH accretion histories of the mergers (see
Fig. \ref{Dist_SFR_BH-evo}). The higher the initial BH masses the lower
the resulting SFRs are throughout the merger. The SFRs are suppressed by a factor of $\sim4$ in the
O-series and a factor of $\sim8$ in the A-series simulations compared to the
B-merger series. In addition, massive BHs suppress the peak BH accretion
rate during final coalescence by a factor of $\gtrsim 10$.

\begin{figure*}
\centering 
\includegraphics[width=14.9cm]{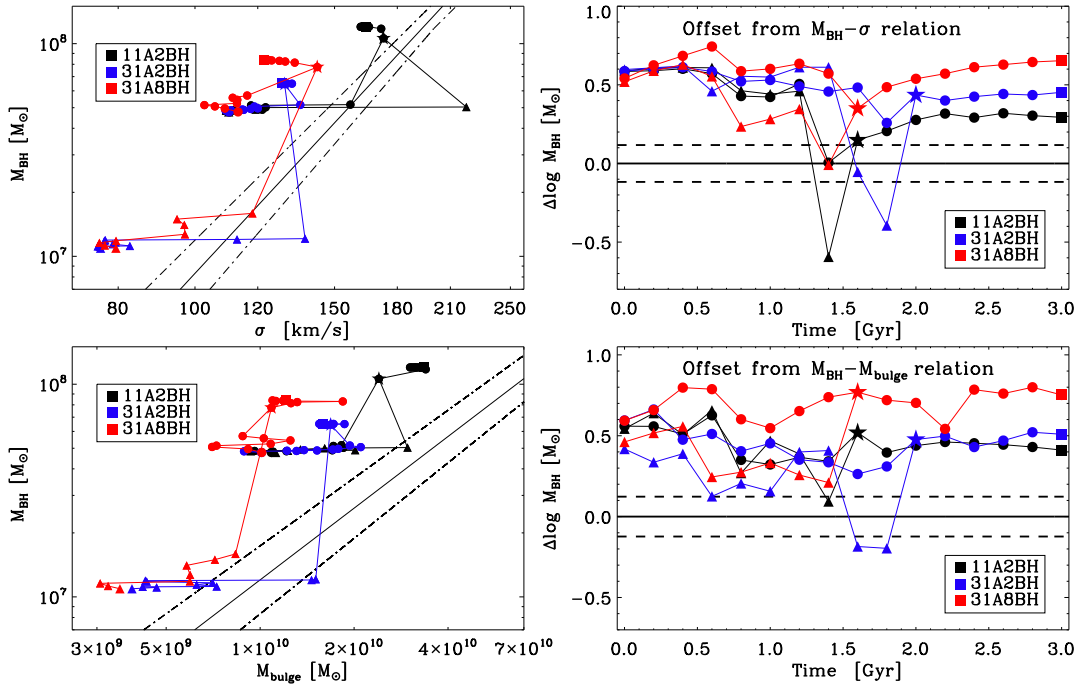}
\caption{The evolution of the A-merger sample with the BHs starting
  initially above the local scaling relations $(\Delta \log M_{\rm
    BH,i}\sim 0.5)$. Symbols and the observed relations
  as defined in Fig. \ref{MBH_evo_new_BBH}.}
\label{MBH_evo_new_ABH}
\end{figure*}

Following observational estimates we calculate from each snapshot the
line-of-sight stellar velocity dispersion $\sigma$ inside the
effective radius using 50 randomly projected realizations of the
galaxies. Together with the extracted BH masses we then plot the evolution of
the BHs during the mergers on the $M_{\rm BH}-\sigma$ plane (top left panels
in Figs. \ref{MBH_evo_new_BBH}-\ref{MBH_evo_new_ABH}). Here we also overplot the
observed local relation ($\log (M_{\rm BH}/M_{\odot})=a+b \log
(\sigma/\sigma_{0})$, $a=8.13\pm0.06$, $b=4.02\pm0.32$,  $\sigma_{0}=200 \ \rm{km
s^{-1}}$, \citealp{2002ApJ...574..740T}) as a solid line with the dashed lines indicating the
$1\sigma$ errors. In the top right-hand panels of
Figs. \ref{MBH_evo_new_BBH}-\ref{MBH_evo_new_ABH} we calculate as a function of time 
the relative offset of the simulated BH masses from the ones predicted by the relation defined as,
\begin{equation}
\Delta \log M_{\rm BH}=\log M_{\rm BH,simulation} - \log M_{\rm BH,predicted}
\label{eq:offset}
\end{equation}
where a positive $\Delta \log M_{\rm BH}$ indicates an overmassive and a
negative  $\Delta \log M_{\rm BH}$ an undermassive BH with respect to the
observed correlation (the $M_{\rm BH}-\sigma$ relation
in this instance). The dashed lines in the plots give the mean $1\sigma$ error
in the velocity dispersion range probed by our simulations.

In the bottom panels of Figs. \ref{MBH_evo_new_BBH}-\ref{MBH_evo_new_ABH} we plot the
evolution of the BHs on the $M_{\rm BH}-M_{\rm bulge}$ plane, with the total bulge masses derived
following the method outlined in \citet{2006MNRAS.369..625N}. Artificial
images smoothed with a Gaussian filter were created of
every snapshot seen from 50 random projections by binning the central 30 kpc
into 128x128 pixels. We then fit both a pure S\'ersic $(\Sigma(r)=\Sigma(0)e^{-b_{n}(r/r_{e})^{1/n}})$
as well as an exponential disk and S\'ersic bulge component simultaneously 
$(\Sigma(r)=\Sigma_{D}(0)e^{r/h_{D}}+\Sigma_{B}(0)e^{-b_{n}(r/r_{eB})^{1/n_{B}}})$. Here
$r_{e},r_{eB}$ are the effective radii and $n,n_{B}$ the S\'ersic indices of the
bulge component, $h_{D}$ is the scale length of the disk, $\Sigma(0)$ are the central
surface densities and $b_{n}$ is chosen so that $r_{e}$ encloses half the
total mass. We then solve for the best fit bulge-to-total (B/T) ratio for each
projection and give the total stellar bulge mass as the mean of the best fit
values. In addition, we overplot the observed local $M_{\rm BH}-M_{\rm bulge}$ relation with errors 
$(\log (M_{\rm BH}/M_{\odot})=c+d \log (M_*/10^{11} M_{\odot})$,
$c=8.20\pm0.10$, $d=1.12\pm0.06$, \citealp{2004ApJ...604L..89H}) together with
the corresponding evolution of $\Delta \log M_{\rm BH}$ as a function of time in the bottom right panels.

\section{Results}

\subsection{Mergers with $M_{\rm BH}$ starting below the relations}
\label{below_rel}

We plot in Fig. \ref{MBH_evo_new_BBH} the evolution of our B-merger sample on the
planes of the two BH scaling relations. For this sample the initial BH mass
is always set to $M_{\rm BH,i}=10^{5} \ M_{\odot}$ $(\Delta \log M_{\rm
    BH,i}\sim -2)$ similarly to J09 and thus rendering the BH initially dynamically unimportant. 

All models show initially relatively modest evolution towards the
observed relations, with the first passage at $t\sim 0.6 \ \rm Gyr$
primarily only seen as a mild increase in $\sigma$ and $M_{\rm bulge}$. 
In the phase $t\sim 0.6-1.2 \ \rm Gyr$ when the galaxies are falling back
towards the second encounter both $\sigma$ and $M_{\rm bulge}$ are
increasing more rapidly, due to stronger merger-induced tidal disturbances. 
After the second encounter the galaxies rapidly merge within
$t\sim 0.2-0.4 \ \rm Gyr$, with the shorter timescales corresponding to the
equal-mass merger and the merger with the higher initial gas fraction of
$f_{\rm gas}=0.8 $ (see Fig. \ref{Dist_SFR_BH-evo}). 

This evolution is qualitatively similar to the early results of \citet{2006MNRAS.372..839N,2006ApJ...651..835D}, who
found that the velocity dispersion increases primarily in two phases, with
mild increase after the first encounter and very rapid evolution at the time
of the coalescence. In \citet{2006ApJ...651..835D} the authors also conjectured that the BHs
would fall on the $M_{\rm BH}-\sigma$ already after the first encounter if 
a constant fraction of 1\% of the central gas supply would be accreted
onto the BH. However, these early studies were purely hydrodynamical, lacking
self-consistent BH growth and feedback.

We find that although the BHs evolve towards the relation during the
merger, the final relations are only established at the time of the final
coalescence of the BHs. Typically the velocity dispersions of the remnants are
somewhat too large immediately after the merger. The effective velocity
dispersion then slowly decreases due to some low-level residual star formation
with the stars preferentially settling in a disk-like structure, with this effect being
stronger in high-gas fraction mergers. However, this effect is less strong in
low-gas fraction and low mass ratio mergers, as star formation and BH mass
growth are more strongly terminated in these mergers after final 
coalescence (Fig. \ref{Dist_SFR_BH-evo}, see also \citealp{2005ApJ...620L..79S,2008AN....329..956J}).   
On the other hand, we find that the bulge mass is established on a short
timescale immediately after coalescence with the pre-coalescence bulge mass
typically being markedly lower.

The prime driver for the scatter is caused by the initial gas fraction $f_{\rm gas}$
of the progenitor galaxies, with the scatter produced by the evolutionary stage of the merger
remnants after the final coalescence being only a secondary effect, as can be seen in the relatively
modest evolution between coalescence (star symbols) and the final merger
remnant state (square symbols).

\subsection{Mergers with $M_{\rm BH}$ starting on the relations}
\label{on_rel}

In Fig. \ref{MBH_evo_new_OBH} we study the evolution of the O-merger sample, for
which we set the initial seed BH mass such that the model galaxies lie on the
observed relations $(\Delta \log M_{\rm BH,i}\sim 0)$. 

Both the primary and secondary galaxies in the mergers typically evolve initially along the observed
relations, but after the first encounter before final coalescence the
secondary galaxy typically evolves horizontally off the relations, again due
to the tidal disturbances after the first passage (Fig. \ref{MBH_evo_new_OBH}). However, after the merging of the BHs all remnants
again evolve rapidly onto the relations, with the exception of the high gas
fraction merger, which evolves off the relation. Thus for progenitors
with initially low gas fractions ($f_{\rm gas}=0.2$) the BH feedback is able
to self-regulate the growth of the BHs maintaining them on the relation. For
very high gas fractions the BH growth is too efficient with the self-regulation
being unable to maintain the galaxies on the observed relations.

\subsection{Mergers with $M_{\rm BH}$ starting above the relations}
\label{above_rel}

Finally, motivated by the observations of
\citet{2006ApJ...645..900W,2008ApJ...681..925W} we run simulations in which
the BHs were initially overmassive by a factor of three $(\Delta \log M_{\rm
  BH}\sim 0.5)$. 

The evolution of this A-merger sample is shown in Fig. \ref{MBH_evo_new_ABH}.
Again we see initially weak evolution with the initial offset maintained, but after
the first encounter the secondary BH evolves horizontally towards the
observed relation and in some instances even below the relations. After the merger
all remnants again evolve above the relations. However, the final merger
remnants of the low gas fraction ($f_{\rm gas}=0.2$) simulations have a 
lower offset $(\Delta \log M_{\rm BH})$ by $\sim0.2$ dex compared to the
initial state. This effect is stronger in the 1:1 merger, which is experiencing a
more violent encounter, with more gas channeled to the center and thus
stronger self-regulation of the BH. The high gas fraction merger ($f_{\rm
  gas}=0.8$) on the other hand evolves even further away from the observed
relations increasing the offset typically by $\sim0.2$ dex.

\section{Discussion}

In this Letter, we have studied the evolution of galaxy mergers on the $M_{\rm
  BH}-\sigma$ and $M_{{\rm BH}}-M_{\rm bulge}$ planes as a function of time. We have shown that 
progenitors with low initial gas fractions ($f_{\rm gas}=0.2$) starting below the relations evolve onto the relations,
progenitors on the relations stay there and progenitors above the
relations evolve towards the relations, but still remaining above
them. Progenitors with high initial gas fractions ($f_{\rm gas}=0.8$) evolve
above the relations in all cases, with the initial gas fraction thus being the
prime source of scatter in the observed relations (see also \citealt{2007ApJ...669...45H}). The evolution for all
mergers is initially slow with the observed relations typically being rapidly
established during a relatively short phase centered at the time of final coalescence of the BHs.

The observations of \citet{2006ApJ...645..900W,2008ApJ...681..925W} indicate 
that the BHs at high redshifts were typically overmassive by a factor of $\sim3$ for a 
fixed $\sigma$ and $M_{\rm bulge}$ compared to the local relation. These BHs
could plausibly have been formed during a brief quasar phase
(e.g. \citealp{2006ApJS..163....1H}) triggered during the mergers of very gas-rich galaxies, thus
resulting in overmassive BH masses for their given $\sigma$ and $M_{\rm
  bulge}$, as seen in our $f_{\rm gas}=0.8$ simulation series. 
However, this being the
case it is not obvious how these galaxies would evolve onto the local observed
relations until the present day. Another binary merger with massive BHs in place does not bring the
galaxies onto the relations (O- and A-series), with high gas fractions mergers moving
them even further away from the relations. Potentially, the bulge mass could
be increased by dry minor mergers (e.g. \citealp{2009ApJ...699L.178N}),
whereas internal secular processes could be
responsible for increasing the velocity dispersion. However, it is not obvious how 
both the velocity dispersion and the bulge mass could be increased at a fixed
BH mass. Finally, another possibility could be that some of the BHs undergoing
mergers are ejected in a sling-shot effect \citep{1974ApJ...190..253S}, thus
decreasing the total BH mass.

The fact that the BHs starting below the relations evolve onto the
relation, whereas the ones above do not, indicates that our simulations
rule out the scenario in which overmassive BHs evolve onto the
relations through gas-rich mergers, whereas undermassive BHs can evolve onto
the relations in galaxy mergers. Thus, given the numerical limitations inherit in our simulations,
our results disfavor the picture in which supermassive BHs develop significantly before their parent bulges.

Finally, our current BH accretion and feedback prescription seems to describe the
growth of BH during the merger phase adequately. However, it is not obvious
that the present description is also valid during the secularly driven low
accretion phase after the merger is completed. Some initial steps to improve
the prescription have been taken by developing an entirely new momentum driven
feedback prescription \citep{2009arXiv0909.2872D}. Recent observations \citep{2007MNRAS.382.1415S,2009ApJ...692L..19S} have indicated
that there is a time lag of $\sim0.5 \ \rm Gyr$ between the peak of star formation
and the onset of AGN activity. Our current model has difficulties in reproducing
this and thus developing models that shed
light on this discrepancy might also ultimately help us understanding in how,
when and why the BHs in galaxies evolve onto the observed relations.

\begin{acknowledgements}

The numerical simulations were performed on the local SGI-Altix
3700 Bx2, which was partly funded by the Cluster of Excellence: ''Origin and
Structure of the Universe''. 

\end{acknowledgements}


\end{document}